# Efficient Light Funneling based on the non-Hermitian Skin Effect

Sebastian Weidemann[1†], Mark Kremer[1†], Tobias Helbig[2], Tobias Hofmann[2], Alexander Stegmaier[2], Martin Greiter[2], Ronny Thomale[2], and Alexander Szameit[1*]

[1]Institut of Physics, University of Rostock, Albert-Einstein-Straße 23, 18059 Rostock, Germany.

[2]Department of Physics and Astronomy, Julius-Maximilians-Universität Würzburg, Am Hubland, 97074 Würzburg, Germany.

*Correspondence to: alexander.szameit@uni-rostock.de

†Those authors contributed equally.

**In the last two decades, the ubiquitous effect of dissipation has proven to entail astonishing non-Hermitian features, rather than just being an inescapable nuisance. As an alternative route to non-Hermiticity, we tailor the anisotropy of a lattice, which constitutes an, up to now, barely exploited degree of freedom. In this case, the appearance of an interface dramatically alters the entire eigenmode spectrum, leading to the exponential localization of all modes at the interface, which goes beyond the expectations for Hermitian systems. This effect is dubbed "non-Hermitian skin effect". We experimentally demonstrate it by studying the propagation of light in a large scale photonic mesh lattice. For arbitrary excitations, we find that light is always transported to the interface, realizing a highly efficient funnel for light.**

As the exchange of energy with the environment is inevitable for any physical system, it becomes clear that non-Hermitian phenomena are ubiquitous in all day live. While they were often considered to be simply undesirable side effects, this perception changed in recent years (*1–5*) when specifically tailored gain and loss distributions lead to intriguing features like non-orthogonal eigenmodes (*6*), exceptional points (*7–9*) and peculiar transport transitions (*10*). By now, already several promising technical applications have been proposed, such as exceptional-point enhanced sensing (*11*), or mode selective laser cavities (*12, 13*). Especially in the field of photonics, the successful integration of optical gain and loss enabled the experimental demonstration of a plethora of promising non-Hermitian features (*4, 14–17*) such as loss-induced transparency (*18*) or unidirectional invisibility (*19, 20*), paving the road for a new generation of optical devices.

Only recently it has been conjectured that in certain lattices where the source of non-Hermiticity does not derive from gain and loss but anisotropic coupling, the introduction of an interface causes all eigenmodes to localize at this interface, independent of coexisting topological boundary states. This phenomenon has recently been dubbed "non-Hermitian skin effect" (*21–25*) and is currently the key element in a lively debate about the validity of the bulk boundary correspondence (BBC) in non-Hermitian topological systems (*21, 22, 26–29*). An independent experimentally viable application of this effect has, however, so far been elusive.

In our work, we experimentally demonstrate the non-Hermitian skin effect for a photonic lattice. That is, we show that the appearance of an interface/boundary in particular non-Hermitian lattices induces a strong and non-local change of the whole spectrum, in contrast to the common wisdom from Hermitian systems that an interface changes the spectrum of eigenstates only locally. Indeed all bulk modes transform into boundary modes in the presence of an interface, rendering the notion of bulk versus edge modes obsolete.

We implement the non-Hermiticity by deliberately breaking the coupling isotropy, and hence reciprocity, of a lattice. To overcome the technical difficulties associated with the realization of a direction-sensitive coupling, we employ a time-bin encoded photonic mesh lattice, which has proven to be a powerful experimental platform especially in the realm of non-Hermitian physics (*7*, *30–32*). In particular, our experiments demonstrate that the dramatic changes in the mode spectrum initiate an intriguing funneling of light. Independently from the excitation position, light is transported to the interface and remains there. This opens up a whole new perspective for novel artificial structures, tailored to collect light in an extremely efficient manner. With this in mind, we further probe the robustness of the non-Hermitian skin effect against phase disorder by studying the impact of Anderson localization. We find a continuous crossover between the skin effect localization at the interface and Anderson localization at the distant excitation site.

When dealing with periodic systems, one often uses the approximation of infinitely expanded settings, which is - of course - never fulfilled in any experiment. This approximation is commonly justified, because in sufficiently large systems, any distant boundaries are expected to introduce only minor and especially local changes in the mode spectrum, hence not causing significant deviations from the infinite system. A prominent example for this approximation, to a generalization of which we will particularize in the following, is the topological boundary mode in the Hermitian Su Schrieffer Heeger (SSH) (*33*) model. Besides the possible appearance of a topological boundary mode due to atomic limit obstruction, the eigenmode spectrum of the bulk does not change significantly upon introducing a boundary.

However, this situation drastically changes when turning to non-Hermitian contributions. Especially for the case of coupling anisotropy in our work, the appearance of a boundary significantly alters the entire mode spectrum in a non-local fashion, as all modes immediately localize at the boundary, also independent of the presence of topological boundary states. This phenomenon has been dubbed non-Hermitian skin effect (*21–25*). As a consequence, and as we will show in our work, in a corresponding photonic system any light signal is transported to that boundary, irrespective of the initial amplitude distribution and position, creating a funnel like behavior, which is schematically shown in Fig. 1A.

We start with our report by illustrating the concept of the non-Hermitian skin effect, and contrast it to the dynamics in Hermitian systems. To this end, we consider a chain of nearest neighbor coupled lattice sites with alternative coupling constants, i.e., the Hermitian SSH model (*33*) displayed in Fig. 1B. Here, every second coupling is chosen to be different but isotropic, indicated by the different shades of the orange arrows. In Fig. 1C, an anisotropic - non-Hermitian - coupling is introduced, such that the hopping from a site to its left neighbor is different from the hopping to its right neighbor.

Due to translational invariance, the eigenmodes in both lattices are delocalized when periodic boundary conditions are applied. When introducing an interface in the SSH lattice by inverting the ratio $c_1/c_2$ at some position (illustrated in Fig. 2A, where the inverted ribbon indicates the inverted coupling ratio), only one mode localizes at the interface, which is well-known as a topological

SSH mode (*34*). All other modes, however, remain delocalized; that is, far away from the interface the modal amplitudes on the individual sites remain essentially unchanged (Fig. 2B).

The situation changes drastically when an interface is introduced in the non-Hermitian lattice with anisotropic hopping. The interface is created by flipping the direction of the anisotropy at some position, illustrated by the mirrored pattern in Fig. 2C. As a consequence, we find that the entire eigenmode spectrum collapses and all eigenmodes are exponentially localized at the interface, as shown in Fig. 2D. In other words, everywhere in the lattice, irrespective of the distance to the interface, the eigenmodes feel the presence of the interface in a non-local fashion. All former bulk modes transform into boundary modes, making the notion of bulk, compared to edge modes, invalid, demonstrating the inapplicability of the BBC. This behavior is in stark contrast to the Hermitian case, where sufficiently far from an interface its influence on the bulk mode structure is negligible. The localization of the eigenmodes in the non-Hermitian lattice has further profound consequences: No matter where the lattice is excited, every signal travels towards the interface. In the context of photonics, this means that any light signal that impinges the lattice is guided towards the interface and remains there. This "non-Hermitian funnel for light" may be the basis for novel intriguing applications for the efficient collection of light.

The theoretical basis of our studies is a modified version of a one-dimensional discrete-time quantum walk, also called light walk (*35*). The dynamics are governed by the evolution equations (1) and (2).

$$u_n^{m+1} = \cos(\beta) u_{n+1}^m + i \sin(\beta) v_{n+1}^m \qquad (1)$$
$$v_n^{m+1} = i \sin(\beta) u_{n-1}^m + \cos(\beta) v_{n-1}^m \qquad (2)$$

Here $u_n^m$ denotes the amplitude at lattice position *n* and time step *m*, on left moving paths, and $v_n^m$ the corresponding amplitude on right moving paths. The parameter $\beta = \beta(n,m)$ characterizes the splitting ratio of the beam splitter between the two paths, where $\beta = \pi/4$ corresponds to a 50:50 coupler. The beam splitter mediates the hopping between lattice sites, as depicted in Fig 3A. Therefore, the hopping between the paths manifests in the splitting ratios of the beam splitters, characterized by $\beta$. In this vein, the SSH model is realized by changing the coupling ratio of every second beam splitter, indicated by the different shades of orange of the arrows and beam splitter cubes. A detailed treatment of the two models discussed in our work can be found in the supplementary material.

In order to experimentally realize the non-Hermitian skin effect, we require an anisotropic hopping that is achieved by the introduction of amplification or attenuation depending on the hopping direction, as shown by the green plus and minus signs, respectively (Fig 3A). This modulation is equivalent to replacing a common beam splitter with its anisotropic counterpart (see supplementary material). Such a modulation was already used in previous works to study PT-symmetric Bloch oscillations (*30*, *36*).

Now we turn to the experimental implementation of our ideas. The light walk is realized by coupled optical fibers, where optical pulses propagate in two unequally long loops, which are connected by a variable beam splitter (VBS) (Fig. 3B). The pulse dynamics in the fiber loop arrangement can be mapped onto a large-scale 1+1D mesh lattice with discrete propagation steps (see supplementary methods). In addition, we employ acousto-optical modulators (AOMs) to manipulate amplitudes and a phase modulator (PM), which are used for the disorder analysis below.

We start our experiment by probing the Hermitian SSH model that is realized by implementing two different coupling ratios $\beta_1$ and $\beta_2$. In order to characterize the transport in the lattice we excite it at three different positions: left from the interface (Fig. 3C), directly at the interface (Fig. 3D), and right from the interface (Fig. 3E). Clearly, any excitation populates extended modes that lead to a spreading of the wave packet, even at the interface. This is consistent with the spectrum of eigenmodes of this lattice shown in Fig. 2B, where only one localized (topological) mode exists at the interface, whereas all other modes remain delocalized and spread over the entire lattice.

The picture changes significantly when probing our non-Hermitian lattice. As all eigenmodes are localized at the interface (as shown in Fig. 2D), any excitation results in a light flow that is directed towards the interface. This is true when exciting left from the interface (Fig. 2F), at the interface (Fig. 2G), and right from the interface (Fig. 2H). This is exactly the manifestation of the non-Hermitian skin effect and the inapplicability of the BBC: The presence of an interface forces the eigenmodes to collapse at the interface, and no delocalized modes remain in the bulk of the lattice. As a consequence, any light excitation somewhere in the lattice exhibits a funnel-like behavior, such that light exclusively localizes along the interface.

In the next set of experiments we probe the robustness of the light funneling effect against disorder. In more physical terms, we explore the interdependence of the non-Hermitian skin effect with Anderson localization (*37*, *38*). This could be important, as Anderson localization is based on long-range interference, which might be suppressed by the non-Hermitian modulation. We address this problem by adding a uniformly distributed phase disorder in space *n* upon the amplitudes $u_n^m$, which is experimentally realized by using the fiber-based phase modulator (see Fig. 3B). In a first step, we combine the anisotropic modulation with a weak disorder (Fig. 4A), where the light is still moving towards the interface, showing the robustness against a certain degree of disorder. If, on the other hand, strong disorder is applied, one can observe that the wave packet localizes and no movement to the interface occurs (Fig. 4B). This behavior is equivalent to disorder induced Anderson localization without anisotropy modulation (*39*, *40*), as shown in Fig. 4C. These results clearly show that the non-Hermitian modulation in our lattice does not inhibit the interference-based effect of Anderson localization, and a continuous crossover between the skin effect localization at the funnel opening and Anderson localization at the excitation site can be found (see Supplementary material). This is in contrast to theoretical works of similar but more complex settings, where the transition from the non-Hermitian skin effect to Anderson modulation undergoes a delocalization phase (*37*).

In our work, we experimentally demonstrated that the bulk eigenmodes of non-Hermitian systems can possess an exceptional sensitivity to the existence of boundaries, that goes far beyond our expectations for Hermitian systems. To this end, we used a sparsely exploited degree of freedom to introduce the non-Hermiticity: The breaking of the isotropy and reciprocity of the underlying lattice. In such a system, the introduction of an interface forces the entire mode spectrum to exponentially localize at the interface, which is the manifestation of the non-Hermitian skin effect. As a result, light that is launched into the lattice is always transported towards the interface, irrespective of the excitation position. Our system, therefore, acts as a funnel for light and allows the efficient collection of any light distribution launched into the lattice. Finally, we also demonstrated that the non-Hermitian skin effect, although being robust to moderate disorder, eventually does not hinder interference-based localization, which we demonstrate by observing a continuous transition to Anderson localization at growing disorder strength. We experimentally implement the theoretical predictions by measuring the propagation of light in a photonic mesh lattice. Those findings may pave the way for utilizing non-Hermitian physics to collect light in a novel and efficient way. Since this approach is not limited to the specific experimental platform and based only on the wave-like properties of light, it might even spark into other areas of research using different experimental environments.

**Acknowledgments:** The authors thank Martin Wimmer for very useful discussions.

**Funding:** The authors acknowledge funding from the Deutsche Forschungsgemeinschaft (BL 574/13-1, SZ 276/19-1, SZ 276/20-1, SZ 276/9-2, and 258499086 - SFB 1170, the Würzburg-Dresden Cluster of Excellence on Complexity and Topology in Quantum Matter *ct.qmat* (39085490 - EXC 2147), and the Alfried Krupp von Bohlen und Halbach Foundation.

**Author contributions:** M.K. developed the theory and S.W. performed the experiments on the photonic mesh lattice. R.T. and A.S. supervised the project. All authors discussed the results and co-wrote the paper. The manuscript reflects the contributions of all authors.

**Competing interests:** Authors declare that they have no competing interests.

**Data and materials availability:** All experimental data and any related experimental background information not mentioned in the text are available from the authors on reasonable request.

**Figures:**

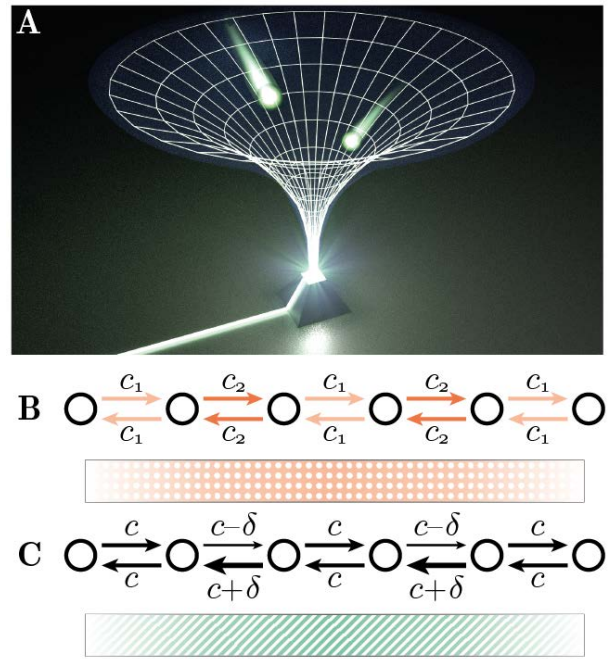

**Fig. 1. General Model** (A) Illustration of the light gathering concept, which reassembles a funnel for light. (B) Linear chain of weakly coupled sites, with the coupling strength $c_1$ and $c_2$. This modulation is identified by the dotted orange ribbon below the chain for reasons of clarity in the other figures. (C) Linear chain of weakly coupled sites, with the coupling strength $c + \delta$ in the left direction and $c - \delta$ in the right direction for every second coupling. This anisotropic modulation is identified by the ribbon with angled green stripes below the chain for reasons of clarity in the other figures.

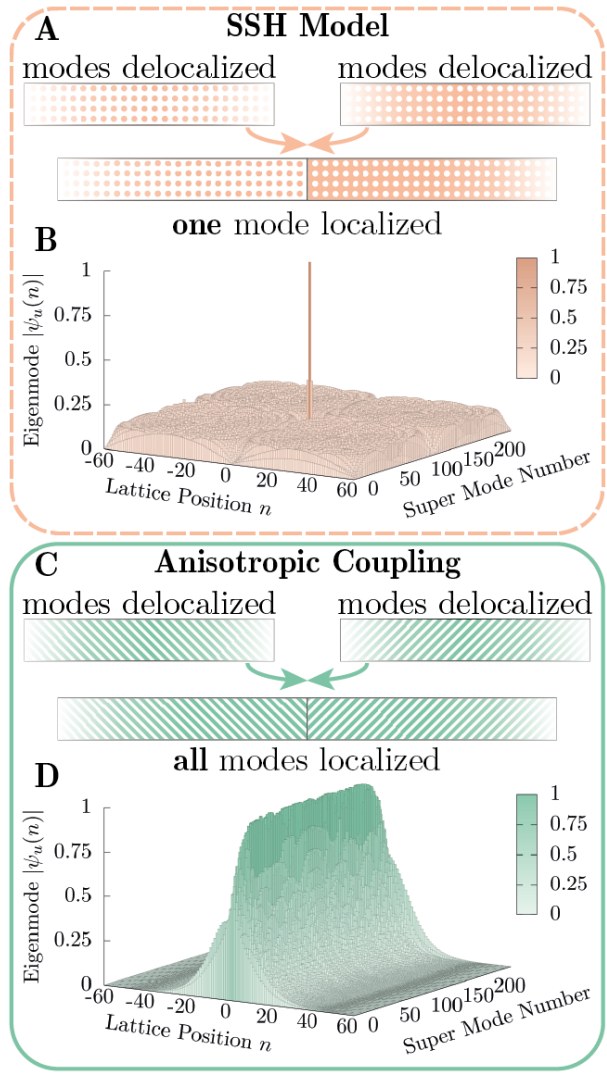

**Fig. 2**. **Eigenmodes of Systems with Interfaces.** (A) An interface is formed with two lattices with SSH modulation, presented in Fig. 1C, where the inverted ribbon indicates the inverted coupling ratio. (B) The eigenmodes of the system with interface, presented in (A) is plotted for a lattice with 120 lattice sites. (C) An interface is formed with two lattices with anisotropic modulation, presented in Fig. 1D, where the inverted ribbon indicates the inverted anisotropy $\delta \to -\delta$. (D) The eigenmodes of the system with interface, presented in (C) is plotted for a lattice with 120 lattice sites.

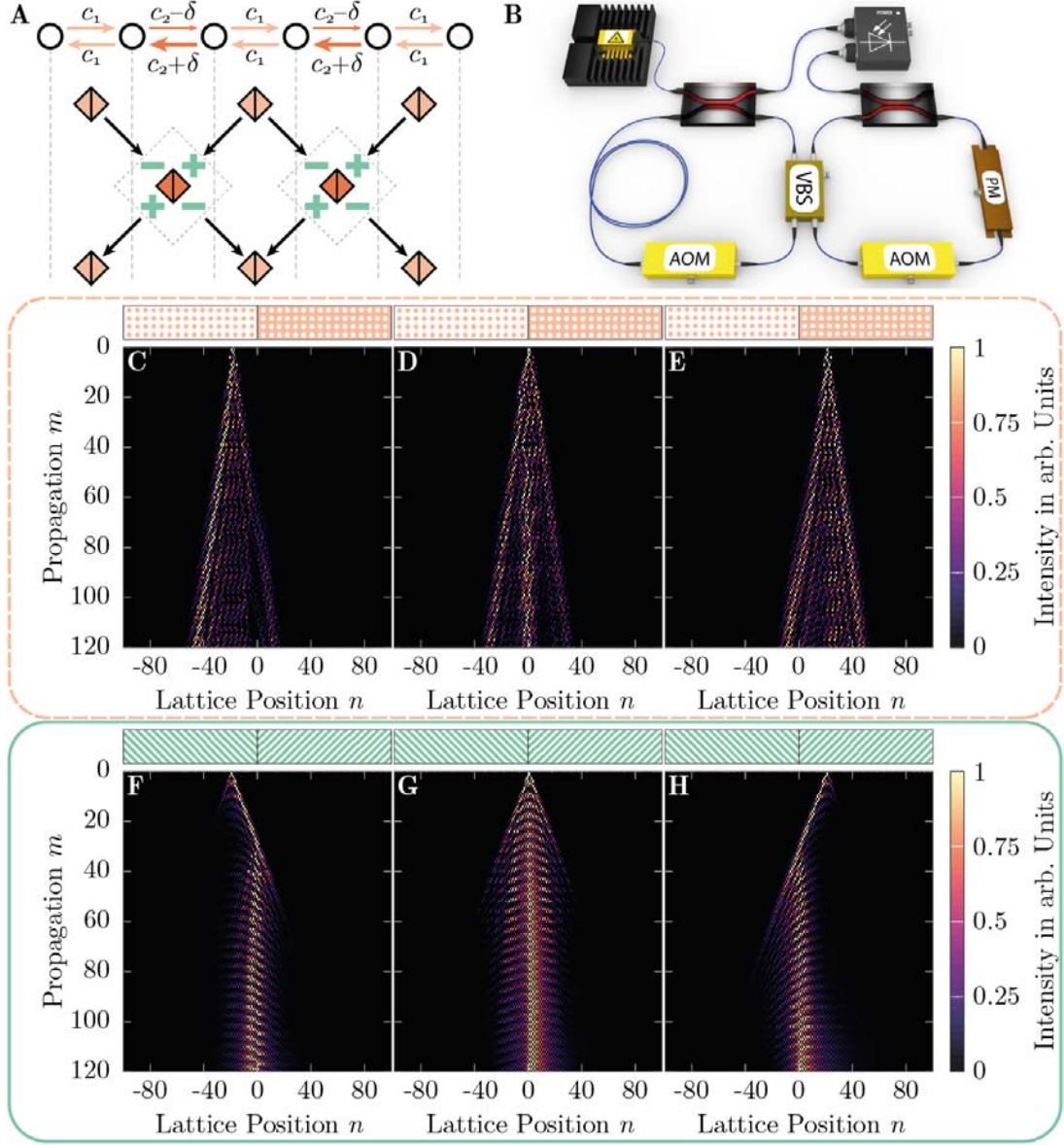

**Fig. 3**. **Experimental Setup and Measurements.** (A) Mapping from the models presented in Fig. 1B and C to a light walk. The different shades of orange represent different coupling strengths (coupling ratios), while the green + plus and – minus signs stand for amplitude modulations. Note that in the experiments either the coupling modulation (SSH) or the amplitude modulation (skin effect) is applied and never both combined. (B) The experimental setup consists of two fiber loops, which are connected by a variable beam splitter (VBS). One loop is connected to a pulsed laser source. The propagation of pulses through the loop arrangement can be mathematically mapped to a propagation through a mesh lattice of beam splitter cubes (A). The amplitudes and phases of the pulses are manipulated by an acusto-optical modulator (AOM) and a phase modulator (PM), respectively. (C-E) Propagation through the photonic lattice with the SSH modulation for three different excitations, which are at the interface, to the left and to right of it. (F-H) Propagation through the photonic lattice with the anisotropic modulation for three different excitations, which are at the interface, to the left of it and to the right of it.

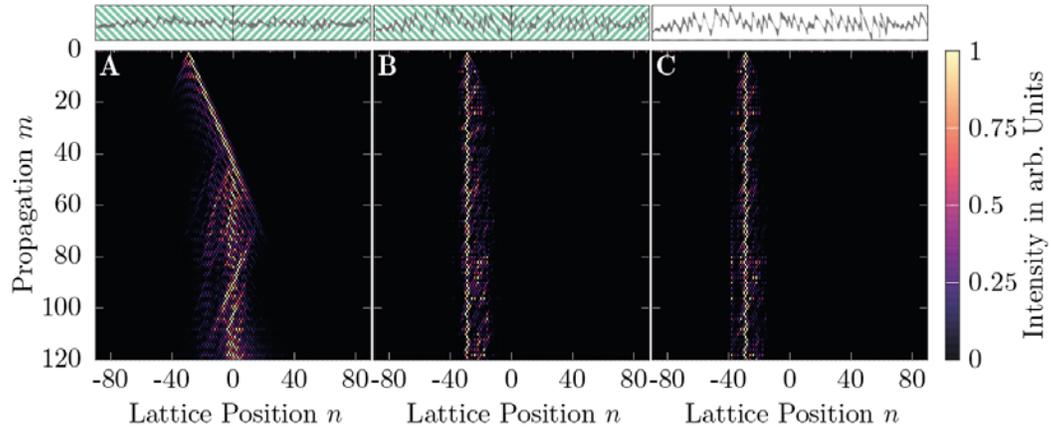

**Fig. 4.** Propagation in the presence of disorder. (A) Propagation in the photonic lattice with the anisotropic modulation and weak phase disorder, with the excitation on the left side of the interface. (B) Same as in (A) but now with strong phase disorder. (C) Same as in (B) but now without the anisotropic modulation.